\documentclass[runningheads]{llncs}
\usepackage[T1]{fontenc}
\usepackage{graphicx}
\usepackage{hyperref}
\usepackage{color}

\usepackage{amsmath}  %
\usepackage[english]{babel}

\usepackage{multirow}

\usepackage{listings}

\definecolor{codegray}{rgb}{0.4,0.4,0.4}
\definecolor{backcolour}{rgb}{0.98,0.98,0.98}
\definecolor{codepurple}{RGB}{127,0,85}
\definecolor{darkolivegreen}{rgb}{0.33,0.42,0.18}

\lstdefinelanguage{sparql}
{
  language=sql, %
  morekeywords={
    prefix,
    filter,
    a,
    inc,
    @prefix,
    ask
  },
  sensitive=false, %
  morecomment=[l]{\#\#}, %
  moredelim=**[is][\color{darkolivegreen}]{@@@}{@@@},  %
  literate=*
    {!=}{{{\color{codepurple}\textbf{!=}}}}2    
    {crm:}{{{\color{darkolivegreen}crm:}}}4
}

\begin{document}

\title{A Three-stage Neuro-symbolic Recommendation Pipeline for Cultural Heritage Knowledge Graphs}
\titlerunning{A Three-stage Neuro-symbolic Recommendation Pipeline for CH KGs}

\author{Krzysztof Kutt\orcidID{0000-0001-5453-9763} \and
El\.{z}bieta~Sroka\orcidID{0000-0001-8952-2187} \and
Oleksandra Ishchuk \and
Luiz do Valle Miranda\orcidID{0000-0003-1838-5693} %
}

\authorrunning{K. Kutt et al.}

\institute{Department of Human-Centered Artificial Intelligence, Institute of Applied Computer Science, Faculty of Physics, Astronomy and Applied Computer Science, Jagiellonian University, prof. Stanis\l{}awa \L{}ojasiewicza 11, 30-348 Krak\'{o}w, Poland\\
\email{\{krzysztof.kutt,elzbieta.sroka,luiz.miranda\}@uj.edu.pl}}

\maketitle              %

\begin{abstract}
The growing volume of digital cultural heritage resources highlights the need for advanced recommendation methods capable of interpreting semantic relationships between heterogeneous data entities.
This paper presents a complete methodology for implementing a hybrid recommendation pipeline integrating knowledge-graph embeddings, approximate nearest-neighbour search, and SPARQL-driven semantic filtering.
The work is evaluated on the JUHMP (Jagiellonian University Heritage Metadata Portal) knowledge graph developed within the CHExRISH project, which at the time of experimentation contained ${\approx}3.2$M RDF triples describing people, events, objects, and historical relations affiliated with the Jagiellonian University (Krak\'{o}w, PL).
We evaluate four embedding families (TransE, ComplEx, ConvE, CompGCN) and perform hyperparameter selection for ComplEx and HNSW.
Then, we present and evaluate the final three-stage neuro-symbolic recommender.
Despite sparse and heterogeneous metadata, the approach produces useful and explainable recommendations,
which were also proven with expert evaluation.

\end{abstract}

\section{Introduction}
\label{sec:intro}

Cultural heritage (CH), including museum and library collections, archives and the achievements of cultural, scientific, and political figures, is an invaluable source of information about the art, history, and social identity of each community.
As a result of mass digitization~\cite{terras2022dh}, the number of digital CH resources increases rapidly,
yet effective exploration remains difficult due to metadata heterogeneity, incompleteness, and the semantic complexity of historical information.
In practice, it means that researchers and other end users often have to juggle several independent search interfaces, fine-tuning queries through the trial-and-error method to achieve satisfactory results.
There is a clear need to develop appropriate tools to help search, explore, and retrieve these vast resources~\cite{li2023towards}.

A promising approach is the development of systems based on semantic technologies---including RDF for data modeling and SPARQL for querying~\cite{schreur2020ld}---which inherently focus on relationships between objects, such as authors, time of creation, function, or place of origin~\cite{europeana2015metadata}.
As a result, all information is modeled as a knowledge graph, which can represent a complex network of connections and contexts~\cite{silva2024ch}.
The approach has been adopted by a variety of cultural institutions, including libraries, museums, archives, and the pan-European aggregator Europeana\footnote{\url{https://www.europeana.eu/}}~\cite{gaitanou2024ld,schreur2020ld,silva2024ch}.
In addition to better metadata management,
the adoption of these technologies has introduced the possibility to use, e.g., semantic search~\cite{chansanam2023ch} or document expansion and query expansion~\cite{almaaitah_opportunities_2020} techniques to facilitate resource searching.
However, they only enhance the standard text- and keyword-based search, without introducing any significant qualitative changes from the user's perspective, so more advanced methods of interaction are still lacking~\cite{li2023towards}.

This paper explores
the feasibility of adapting recommendation systems (RS) to address this need.
RSs are successfully used in many areas such as entertainment, e-commerce, etc~\cite{raza2026rs}.
However, in the field of cultural heritage, such systems are still rare and often limited to simple thematic connections~\cite{casillo2023context,li2023towards}.
In this paper, we discuss what makes CH data unique and why it is so challenging to process.
Then, we propose a three-step recommendation pipeline and evaluate it on a heterogeneous knowledge graph describing the heritage of Jagiellonian University, developed as part of the ongoing CHExRISH project.
Nevertheless, the proposed method is universal and can be applied to a variety of CH datasets.

The rest of the paper is structured as follows.
Related works are summarized in Sect.~\ref{sec:related}.
The proposed pipeline is outlined in Sect.~\ref{sec:methods}
and evaluated in Sect.~\ref{sec:results}.
Sect.~\ref{sec:summary} concludes the paper.

\section{Related works}
\label{sec:related}

\subsection{Cultural heritage knowledge graphs in practice}
\label{sec:ch_data}

Knowledge graphs (KG) are widely used in cultural heritage due to their ability to address the primary challenges of the digital transformation of cultural institutions~\cite{schreur2020ld}, i.e., the need for flexible and interoperable metadata~\cite{almaaitah_opportunities_2020,Haslhofer2018,hyvonen2019knowledge,schreur2020ld}.
Recent mature KG deployments include
\emph{the Europeana} portal, a European Union-funded aggregator that collects, integrates, and offers access to data from ${\approx}3000$ European cultural institutions~\cite{europeana2015metadata},
and \emph{ArCo}, the Italian cultural heritage knowledge graph that contains 169M triples describing 820k entities from MiBAC (General Catalog of the Italian Ministry of Cultural Heritage and Activities)~\cite{carriero2019arco}.
All these efforts are facilitated by a range of diverse standards and dictionaries, including 
\emph{Dublin Core} (\url{https://www.dublincore.org/}) -- one of the more generic vocabularies for Linked Data,
\emph{Records in Contexts Ontology} (RiC-O), %
which enables the representation of archival information within a broader context~\cite{ric2017primer},
\emph{LIDO} (\url{https://lido-schema.org/}), designed to store descriptions of CH objects, related events, and administrative information,
and \emph{CIDOC-CRM}, %
an international standard (ISO 21127:2014) for CH data modeling and sharing across different institutions~\cite{cidoc2014primer}.

Digital metadata on cultural heritage has its own unique characteristics. Key points from the perspective of the reported work:
\begin{enumerate}
    \item The fragmentation and low quality of CH metadata. The analysis of the Task Force on Metadata Quality report shows that the set of 258,494 unique records (\emph{rdf:about}) contains only 61,169 different titles (\emph{dc:title}) and only 11 different descriptions (\emph{dc:description}), indicating the lack of consistent descriptions and the minimum quality required~\cite{europeana2015metadata}.
    \item Limited metadata for objects. Many objects only have rudimentary catalog descriptions, and creating new metadata is a major challenge as it often requires the use of sources that are hundreds or even thousands of years old.
    \item The range of metadata standards used in libraries, museums and archives require multi-stage mapping to a common RDF model, including quality validation and record enrichment, which is associated with time-consuming processes and the need for specialist knowledge~\cite{europeana2015metadata}.
    \item The heterogeneity of metadata from different sources. It is visible especially in aggregators like Europeana~\cite{isaac2012enriching}.
    \item No user history. From the point of view of recommendations, it is very important to note that CH data are made available in publicly accessible systems, without the need to create accounts, and therefore there are no browsing history data sets that could be used for Collaborative Filtering methods.
\end{enumerate}

\subsection{Recommendation systems not only for cultural heritage data}
\label{sec:rs}

Recommendation systems are applications that are used to analyze user preferences and propose the most relevant data.
Such systems can be found in various every-day scenarios, e.g., social networks suggest content to view, e-commerce sites recommend products that users are likely to enjoy~\cite{luyen2023constraint}.
However, traditional recommendation methods, such as collaborative filtering (CF), content-based or demographic approaches, encounter a number of limitations, including the problem of ``cold start'' of new users or products, data sparsity, lack of deeper contextual analysis, and over-specialization~\cite{heitmann2010linked}.
As a consequence, recent surveys~\cite{chicaiza2021survey,zhao2023embedding} document a sustained shift from ``pure'' CF/content-based methods to knowledge-aware recommenders where KGs mitigate the challenges mentioned above while improving interpretability.
In practice, they are typically based on RDF graph data and can use explicitly declared user preferences together with rich and organized domain knowledge.
All of this information is then embedded in a vector space and is used to generate recommendations for records and entities~\cite{zhao2023embedding}.

Key KG embedding families include translation-based (TransE), bilinear (DistMult/ComplEx/RESCAL), convolutional (ConvE), and graph-convolutional (CompGCN/KGAT) models~\cite{zhao2023embedding}.
Recent advances in this field include the strengthening of the classic KG embeddings and relational GNNs, usually paired with the ANN search for efficient generation of candidates for k-NN.
HNSW and FAISS can be selected as a search tool depending on scale-latency-memory constraints~\cite{rahman2024faiss,shah2025hnsw}.
Given curatorial and public-facing requirements for explainability, neuro-symbolic approaches that inject first-order rules/ontologies into representation learning are gaining traction.
Empirical results show gains in accuracy and novelty when rules inform embeddings~\cite{spillo2024rs}.
In the heritage context, curatorial ``paths'' over KGs can act as an interpretation layer above the raw data and the recommendation engine~\cite{mulholland2024supporting}.

Despite the difficulties in modeling cultural knowledge and the insufficient quantity and quality of metadata~\cite{casillo2023context,li2023towards}, several attempts have been made to create recommendation systems for cultural heritage.
\cite{li2023towards} present a method for using rich semantic metadata and graph database capabilities to recommend related exhibits on cultural portals.
The SMARTMUSEUM system~\cite{ruotsalo2013smart} demonstrates a mobile approach to recommendation based on RDF representations of objects, combining contextual information with semantic ranking to provide the user with personalized excursions through the museum.
Finally, the only example of a highly available recommendation system with a large amount of CH data is Europeana's Recommendation API.
It generates personalized suggestions by analyzing relationships among items, entities, or user-curated galleries\footnote{\url{https://pro.europeana.eu/page/the-recommendation-system}}.

\section{Methods}
\label{sec:methods}

\subsection{The JUHMP knowledge graph and CHExRISH context}
\label{sec:chex}

The CHExRISH project\footnote{\url{https://chexrish.id.uj.edu.pl/}}, or Cultural Heritage Exploration and Retrieval with Intelligent Systems, carried out at the Jagiellonian University (JU) in Krak\'{o}w, is an undertaking which, for the first time in the history of our university, brings together University Archives (AUJ), Museum (MUJ), Jagiellonian Library (JL), and teams of IT specialists and humanists in a common mission: to build a new way of accessing our heritage. %
The objectives of these interdisciplinary efforts include: 
integrating and analyzing multiple databases that have been scattered until now,
providing access to AI-based research tools,
and developing best practices for acquiring and protecting heritage data,

One of the main outcomes of the CHExRISH project will be 
a prototype of the JUHMP portal, a shared metadata and research infrastructure for the JU heritage.
The technological basis are knowledge graphs, which will ensure data integration and interoperability. 
As a data structure, graphs enable computational processing that generates new knowledge in the humanities, providing opportunities for network analysis, social network analysis, and recommendations.
Knowledge graphs encode semantic relationships between objects, not only what the objects are, but also how they relate to each other.
This solution has already proven itself in many international projects, such as Europeana, Yale Knowledge Graph, and the Dutch KB Labs.
All of them show that combining museum, library, and archival collections into a coherent network opens up completely new perspectives for researchers, curators, and the general public.
The knowledge graph is to play a similar role in the CHExRISH project
by mapping data from AUJ, MUJ, and JL into a semantic model based on CIDOC-CRM~\cite{lvm2025juhmpprototype1}.
The final CHExRISH graph contains a series of entities following several CIDOC-CRM classes, such as Person, Place, Event, Type etc.

The basis for the unification of the data between the JU units is the matching of personal identifiers between
(1) AUJ's Corpus Academicum Cracoviense (CAC)\footnote{\url{https://cac.historia.uj.edu.pl/}}, an electronic database with around 67,000 records on students and graduates of the University of Krak\'{o}w during the period 1364--1780,
(2) ALMA, an integrated system for the management of around 9 million MARC21-based bibliographic records of JL
and (iii) MuzUJ, a database containing information about museum objects stored at MUJ, containing around 75,000 records.
The information included in the prototypes are the records from each system related to the matched persons.

For the first prototype, a manually created authority file was created~\cite{lvm2025juhmpprototype1} containing 10 individuals. For the second prototype, a random forest model was used using a combination of string similarity metrics to find possible matches~\cite{miranda_wikidata_er_es2c2026}.
The matches were further manually validated by domain experts from AUJ and JL.
A final authority file containing 824 verified matched entities was achieved.
Two versions of the second prototype were used as the basis of the work reported in this paper: \emph{CHExRISH\_Prototype2} (402 948 triples) and \emph{CHExRISH\_FullCAC\_260128} (3 202 711 triples). The latter is an extension of the former including all persons from CAC and their events.

\subsection{Experimental setups}
\label{sec:setup}

Three experimental setups were used in the reported work:
\begin{itemize}
    \item \emph{Laptop} (only CPU): 16 cores, Intel Core U7-165H @ 3.80GHz, 32 GB RAM running on Ubuntu 24.04.3 LTS,
    \item \emph{Ares} (only CPU): 48 cores, 2x Intel Xeon Platinum 8268 @ 2.90GHz, 48 GB RAM running under SLURM queue system\footnote{\url{https://guide.plgrid.pl/en/resources/supercomputers/ares}},
    \item \emph{Local computing server (LCS)} (only GPU): 1x NVIDIA RTX A5500 with 24 GB RAM (CUDA 12.4) running on Ubuntu 22.04.5 LTS.
\end{itemize}

Unless otherwise specified, the tasks were performed on the \emph{Laptop}  to confirm that they are not computationally intensive,
that a CPU-only setup is sufficient for them,
and that they can be easily adapted to other CH projects. %

\subsection{Three-stage recommendation pipeline}
\label{sec:pipeline}

In response to the gaps and challenges in the area of cultural heritage identified above, we have developed our original neuro-symbolic recommendation pipeline, tailored to the specifics of this domain~\cite{ishchuk2025msc}.
It consists of three stages:
\begin{enumerate}
\item \emph{Embedding} trained with PyKEEN~\cite{ali2021pykeen} provides a representation of the RDF graph in a vector space. 
\item \emph{Nearest-neighbor retrieval} using the HNSW index results in fast and efficient candidate generation.
\item \emph{SPARQL-driven semantic filtering} limits the resulting candidates set to the meaningful ones (i.e., provides ontology-aware results) and provides transparent justifications for recommendations
(e.g., same\_objects\_connection, target\_connection).
\end{enumerate} 

The complete code needed to replicate the entire pipeline is available in the repository: \url{https://gitlab.geist.re/pro/chex-recs}.
Input knowledge graphs and intermediate files (e.g., trained embeddings) are available upon request.

\subsection{Embedding model selection and hyperparameters}
\label{sec:embedding}

To determine the most suitable embedding model for the recommendation system, an initial comparative experiment was conducted using four knowledge graph embedding models: TransE, ComplEx, ConvE, and CompGCN.
Each model was trained on the \emph{CHExRISH\_Prototype2} graph for a fixed number of 10 epochs, using identical hyperparameters across all models. This approach ensured that the results remained comparable and were not biased by differences in configuration.

The performance of each model was evaluated using four knowledge-graph completion metrics: \emph{Mean Reciprocal Rank (MRR)}, \emph{Hits@1}, \emph{Hits@3} and \emph{Hits@10}. Additionally, two resource metrics were used: total training time (in seconds) and increase in RAM usage (in megabytes).
As detailed in Tab~\ref{tab:emb_selection}, ComplEx achieved the highest ranking quality, completing 10 epochs in ${\approx}46$ min and increasing RAM by only 64 MB. TransE delivered moderate performance. CompGCN achieved a score higher than ComplEx, but at the cost of an impractical (${\approx}13.3$h) training time. ConvE performed poorly on all quality metrics and consumed the most memory, making it unsuitable for the setup.

\begin{table}[b]
\caption{Embedding models comparison: performance and resource consumption.}\label{tab:emb_selection}
\centering
\begin{tabular}{|l|c|c|c|c|c|c|}
\hline
\emph{Model} & \emph{MRR} & \emph{Hits@1} & \emph{Hits@3} & \emph{Hits@10} & \emph{Time (s)} & \emph{RAM (MB)} \\
\hline
\textbf{TransE} & 0.2868 & 0.2027 & 0.33067 & 0.4352 & 1775.9914 & 161.8203 \\
\textbf{ComplEx} & 0.3515 & 0.2848 & 0.3591 & 0.4763 & 2769.6728 & 63.6445 \\
\textbf{ConvE} & 0.0123 & 0.0012 & 0.0020 & 0.0081 & 2072.7288 & 6486.0156 \\
\textbf{CompGCN} & 0.3132 & 0.1785 & 0.3829 & 0.5594 & 47862.5388 & 64.6758 \\
\hline
\end{tabular}
\end{table}

After selecting the embedding model, it was essential to identify an optimal combination of hyperparameters that would maximize the predictive quality of the ComplEx model while remaining within the available computational resources. An experiment was carried out to assess the impact of two key hyperparameters, the learning rate (\emph{lr}) and the embedding dimensionality (\emph{dim}), on the performance and resource consumption of the ComplEx model. Batch size %
was set to 128,
as it offers a good trade-off on our CPU nodes, fitting in memory, and enabling efficient computations for ComplEx.
The selected results are summarized in Tab.~\ref{tab:hyperparams}.
Overall, the configuration $lr = 0.01$, $dim = 200$ delivered the best balance of accuracy (highest MRR and Hits@k) and acceptable resource requirements.

\begin{table}[t]
\caption{Selected ComplEx parameters comparison: performance and resource consumption.}\label{tab:hyperparams}
\centering
\begin{tabular}{|c|c|c|c|c|c|c|c|}
\hline
\emph{lr} & \emph{dim} & \emph{MRR} & \emph{Hits@1} & \emph{Hits@3} & \emph{Hits@10} & \emph{Time (s)} & \emph{RAM (MB)} \\
\hline
\textbf{0.001} & \textbf{100} & 0.0075 & 0.0027 & 0.0059 & 0.0140 & 830.9784 & 7826.0508 \\
\textbf{0.001} & \textbf{200} & 0.0079 & 0.0026 & 0.0063 & 0.0165 & 1517.0887 & 10752.8516 \\
\textbf{0.010} & \textbf{100} & 0.2011 & 0.1484 & 0.2222 & 0.2952 & 1191.7680 & 10558.1445 \\
\textbf{0.010} & \textbf{200} & 0.2087 & 0.1530 & 0.2286 & 0.3145 & 2146.6507 & 9658.3672 \\
\hline
\end{tabular}
\end{table}

\subsection{HNSW parameters comparison}
\label{sec:hnsw}

The second stage of the proposed pipeline converts the learned ComplEx embeddings into fast nearest-neighbor lookups using the HNSW (Hierarchical Navigable Small World) index, which is implemented with HNSWLib Python module.
The goal is to retrieve, for any query entity, a relatively small set of the most similar entities that can be refined later using SPARQL-based filters.

Before executing the HNSW indexing, a grid-search was performed over key HNSW hyperparameters (graph connectivity: $M \in \{8, 16, 32\}$,
build-time search breadth: $efConstruction \in \{100, 200, 400\}$,
query-time search breadth: $efSearch \in \{50, 100, 150, 200, 300\}$) to determine the best HNSW settings, balancing accuracy (\emph{Recall@k}) and efficiency (\emph{latency}, \emph{memory}, \emph{build time}).
Based on the performed search (see Tab.~\ref{tab:hnsw}), the following configuration was selected: $M = 16, efConstruction = 400, efSearch = 50$. Since index construction is a one-time step and the built index can be subsequently reused from file, build time and memory consumption were not treated as selection criteria. The selection of HNSW hyperparameters was based on retrieval quality and query-time efficiency (\emph{Recall@k} and \emph{mean latency}, respectively).

\begin{table}[t]
\caption{HNSW parameters comparison: recall, latency and resource consumption.}\label{tab:hnsw}
\centering
\begin{tabular}{|c|c|c|c|c|c|c|}
\hline
\emph{M} & \emph{efConstruction} & \emph{efSearch} & \emph{Mean latency (s)} & \emph{Recall@k} & \emph{Time (s)} & \emph{RAM (MB)} \\
\hline
\textbf{16} & \textbf{400} & \textbf{50} & 0.2244 & 1.0 & 6.6428 & 91.3945 \\
\textbf{32} & \textbf{400} & \textbf{50} & 0.2486 & 1.0 & 6.6289 & 97.4023 \\
\textbf{16} & \textbf{200} & \textbf{100} & 0.3369 & 1.0 & 3.3972 & 91.3984 \\
\textbf{16} & \textbf{400} & \textbf{100} & 0.3378 & 1.0 & 6.3348 & 96.5977 \\
\textbf{8} & \textbf{400} & \textbf{150} & 0.3421 & 1.0 & 4.3496 & 88.3984 \\
\textbf{32} & \textbf{100} & \textbf{100} & 0.3541 & 1.0 & 2.7523 & 97.4063 \\
\textbf{32} & \textbf{400} & \textbf{100} & 0.3886 & 1.0 & 7.1688 & 97.4023 \\
\textbf{32} & \textbf{200} & \textbf{100} & 0.3932 & 1.0 & 3.9377 & 97.4023 \\
\textbf{8} & \textbf{100} & \textbf{200} & 0.4136 & 1.0 & 1.6410 & 88.4023 \\
\textbf{16} & \textbf{400} & \textbf{150} & 0.4600 & 1.0 & 6.2475 & 91.4375 \\
\hline
\end{tabular}
\end{table}

\subsection{SPARQL-driven semantic filtering}
\label{sec:sparql}

The final step of the proposed recommendation method refines the embedding-based candidates returned with HNSW by enforcing knowledge-level constraints.
It focuses on the selection and explanation of why a candidate is relevant for specific target, producing machine-readable evidence (lists of matching types, predicates, shared nodes, dates, places) with a set of SPARQL queries.
Semantic filtering stage is a general concept that can be applied in many scenarios, but it relies heavily on the knowledge graph used as input.
Therefore, the SPARQL queries we use are tightly coupled with the structure of the JUHMP knowledge graph and the CIDOC-CRM ontology on which it is based.

As the JUHMP ontology was built primarily around a list of people affiliated with JU, the recommendation method is person-centric. Therefore, the semantic filters constrain candidates to instances of the CIDOC CRM class \emph{crm:E39\_Actor}, that covers human actors, both as single persons and as groups, and its subclass \emph{crm:E21\_Person}, that refers to human beings, either alive or presumed to have lived in the past.%

Then, a predefined set of SPARQL tests is used to check whether candidates are related to the target and how.
Various connections are possible, but limited to a predefined list that includes:
same objects connection,
same location,
same events,
same identification,
same production,
same occurs in,
close birth dates, %
close death dates
and same place of birth.
For example, the ``same events'' test is based on the SPARQL query (see Listing~\ref{lst:sparql}) used to validate whether the target and candidates share the same value in one of the properties:
\emph{crm:P11i\_participated\_in},
\emph{crm:P12i\_was\_present\_at},
\emph{crm:P14i\_performed}.
Only candidates that share at least one exact connection with the target are kept.
As a result, only connected candidates are retained, together with the list of predicates and shared nodes, which supports explanations such as ``both refer to the same event or place''.
For more details, consult 
\emph{04\_generate\_recommendations.ipynb} notebook in the repository.

\begin{lstlisting}[
  language=sparql,
  caption={SPARQL query used to check whether the target and candidates share the same value in given set of properties},
  label=lst:sparql,
  float=b,
]
PREFIX crm: <http://www.cidoc-crm.org/cidoc-crm/>

SELECT DISTINCT ?candidate ?prop ?value
WHERE {
    VALUES ?prop      { {PROPERTIES_URIS} }
    VALUES ?candidate { {CANDIDATES_URIS} }
    <{TARGET_URI}> ?prop ?value .
    ?candidate     ?prop ?value .
}}
ORDER BY ?candidate ?prop ?value
\end{lstlisting}

In summary, the SPARQL-driven filtering transforms vector space neighbors into knowledge-consistent recommendations.
By validating types, checking graph connections, and enforcing domain-specific constraints on time and place, this stage retains only candidates that are semantically plausible and explains each selection with explicit evidence obtained from the graph.
The method has one key practical limitation.
It works best when the graph contains enough structured links (e.g., time spans and places).
If the metadata are sparse, fewer candidates survive the semantic checks.

\subsection{Evaluation approach}
\label{sec:eval_approach}

After selecting models and their hyperparameters (see Sect.~\ref{sec:embedding}-\ref{sec:hnsw}), the recommendation pipeline (see Sect.~\ref{sec:pipeline}) was run on both graphs (see Sect.~\ref{sec:chex}) to evaluate its usefulness.
Embedding for \emph{CHExRISH\_Prototype2} was trained with \emph{Ares} (CPU), while embedding for \emph{CHExRISH\_FullCAC\_260128} was trained with \emph{LCS} (GPU\footnote{This embedding can also be trained in a CPU-only environment, but due to the size of the graph, training on \emph{Ares} or \emph{Laptop} would take over a week. Training on \emph{LCS} took ${\approx}70$h.}).
All other steps were performed on \emph{Laptop}.

The accuracy of the recommendation pipeline was evaluated based on nine scenarios involving candidate generation for nine targets---notable persons from the history of Jagiellonian University (e.g., Nicolaus Copernicus; see Tab.~\ref{tab:experts}).
These people were selected by experts from AUJ, MUJ, and JL.
This approach, focusing on well-known individuals, facilitated subsequent evaluation by experts.

Both trained recommendation models were evaluated in two ways.
Quantitative evaluation focused on verifying basic model performance metrics such as \emph{Mean Reciprocal Rank (MRR)} and \emph{Hits@K}.
The purpose of the expert evaluation was to assess whether the proposed recommendations made sense in terms of content and whether the candidates were indeed historically related to the targets.

\begin{figure}[thp]
    \centering
    \includegraphics[width=.9\textwidth]{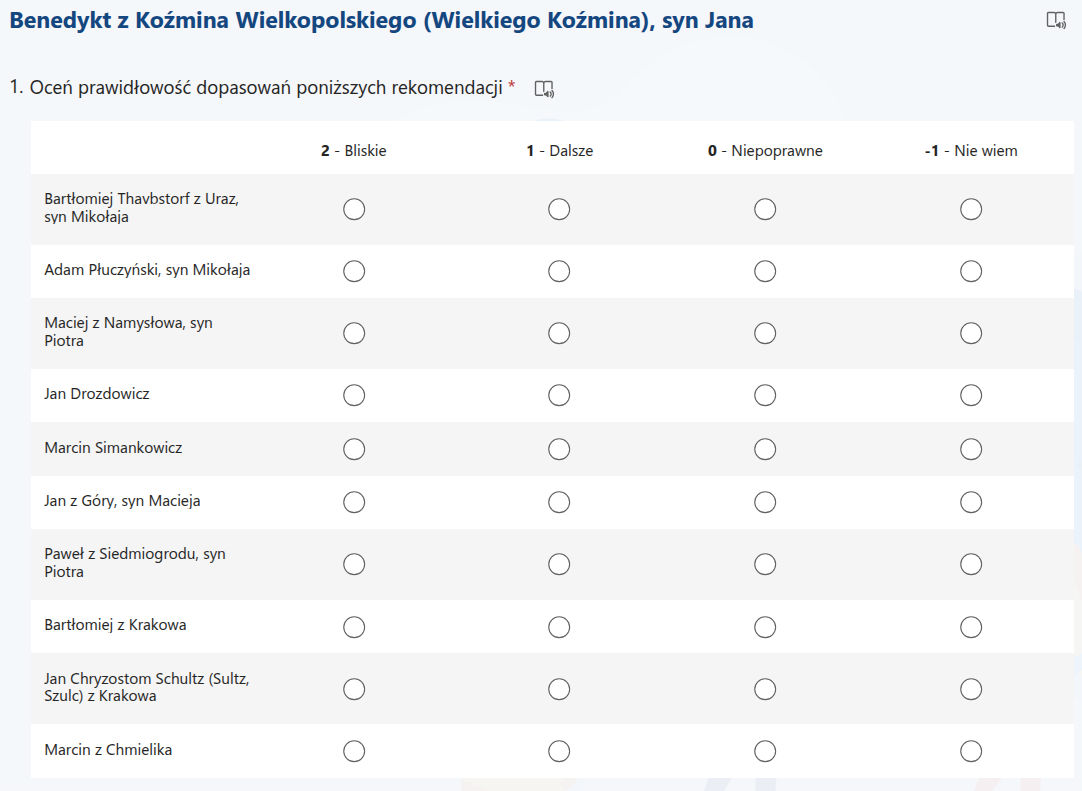}
    \caption{Expert questionnaire. Question about target Benedykt from Koźmin Wielkopolski with the top 10 results generated by the pipeline. The experts were asked to assess the correctness of all 10 results with $[2, 1, 0, -1]$ scale described in the text.}
    \label{fig:form}
\end{figure}

For the expert evaluation, a simple questionnaire was prepared in MS Forms (see Fig.~\ref{fig:form}) and distributed among experts.
For each of the nine scenarios used in the evaluation, the top 10 results generated by the pipeline were selected. Experts were asked to rate the accuracy of the recommendations using a scale:
\begin{itemize}
    \item \emph{2 -- close connection}, a user reading information about person X will be interested in person Y, e.g., people who collaborated with each other, people working on the same research topic, Y was a researcher of X's work, etc.
    \item \emph{1 -- distant connection}, e.g., people lived at the same time but did not collaborate with each other (the user may be interested in such a suggestion)
    \item \emph{0 -- very distant connection (incorrect)}, i.e.: people lived in different times, did not know each other, were involved in different areas of research, etc. (the user will not be interested in such a suggestion)
    \item \emph{-1 -- I do not know this person}.
\end{itemize}
The experts could also add notes and comments, which were not mandatory.

The expert study group included people from AUJ, MUJ and JL. These are people with extensive experience in the field of history and heritage of Jagiellonian University.
Four experts (two responses from AUJ and one each from MUJ and JL) were involved in the evaluation of \emph{CHExRISH\_Prototype2}.
The final model (\emph{CHExRISH\_FullCAC\_260128}) was evaluated by 10 experts: 5 people from the AUJ, 1 from the MUJ, 2 from the JL, 2 did not indicate their unit.

\section{Results and Discussion}
\label{sec:results}

\subsection{Quantitative evaluation}

The final 120-epoch training run of the ComplEx model for \emph{CHExRISH\_Prototype2} resulted with $MRR = 0.2207$, indicating that the correct entity was retrieved within the first four to five positions in the ranking, though not consistently at the very top.
The $Hits@1 = 0.1659$ shows that the correct entity appeared first in only about 16.6\% of test queries, confirming the model's limited ability to rank the correct answer in the leading position. By contrast, the $Hits@3 = 0.2421$ %
and $Hits@10 = 0.3288$ demonstrates that in about one third of the queries the correct entity was found within the first ten positions.
While model consistently surfaces relevant entities within the top 10 results, the relatively low MRR and Hits@1 values indicate difficulty in prioritizing the correct entity at the very top of the ranking.
In terms of resource usage, training the model for 120 epochs required approx. 22 hours on a CPU-only node (\emph{Ares}).

Training for \emph{CHExRISH\_FullCAC\_260128} was completed in 108 hours with $MRR = 0.1393$, $Hits@1 = 0.1057$, $Hits@3 = 0.1572$ and $Hits@10 = 0.2098$.
The metrics in this case are lower than for the small graph, but this graph is also more sparse (there are many people with very limited connections to other nodes in the graph) and more ``dense'' (there were definitely far more people, because instead of just a few prominent figures, this graph included all individuals associated with the University between the 14th and 18th centuries) and therefore more challenging to train.

However, the higher ``density'' of the graph resulted in the pipeline recommending nodes that were generally closer to targets in the embedding space: the similarity of the first recommendation was in the range $[0.40; 0.85]$ for the larger model compared to $[-0.41; 0.21]$ for \emph{CHExRISH\_Prototype2} (cosine similarity ranging from $-1$ to $1$).
Also, the historical completeness of the bigger graph enabled the pipeline to recommend any individuals associated with the University's history in covered time period, so it was assumed that the expert evaluation results would be better, despite the lower quantitative metrics.

\subsection{Expert evaluation}

\begin{table}[t]
\caption{Number of close and distant recommendations as indicated by experts for both models. Max sum for each row in \emph{Prototype2} is 40 (when all 4 experts indicated all 10 recommendations as close/distant). Max sum for each row in \emph{FullCAC\_260128} is 100 (10 experts x 10 recommendations).  Target labels are abbreviated for clarity.}\label{tab:experts}
\centering
\begin{tabular}{|p{5cm}|c|c|c|c|}
\hline
\multirow{2}{5cm}{\emph{Target}} & \multicolumn{2}{|c|}{\emph{Prototype2}} & \multicolumn{2}{|c|}{\emph{FullCAC\_260128}} \\
 & \emph{Close} & \emph{Distant} & \emph{Close} & \emph{Distant} \\
\hline
Benedykt z Koźmina \ldots & 1 & 5 & 2 & 1 \\
Sebastian Sierakowski (hrabia) & 0 & 7 & 1 & 9 \\
Stanisław Reszka z Buku \ldots & 0 & 2 & 2 & 1 \\
Jan III Sobieski (król Polski ; \ldots) & 0 & 1 & 0 & 6 \\
Maciej Karpiga z Miechowa \ldots & 0 & 5 & 2 & 10 \\
Mikołaj Kopernik (Copernicus) \ldots & 0 & 3 & 0 & 6 \\
Jakub Górski (młodszy) \ldots & 0 & 2 & 3 & 6 \\
Andrzej z Buku (starszy) & 0 & 1 & 2 & 6 \\
Jan Brożek (Broscius) \ldots & 0 & 2 & 2 & 6 \\
\hline
\end{tabular}
\end{table}

The experts indicated closer and more distant recommendations proposed by the models for a selected target %
(see Tab.~\ref{tab:experts}).

The results for \emph{CHExRISH\_Prototype2} indicate that of the ten recommendations selected by the model, there are some that agree and have been confirmed by experts during the study, which proves their correctness.
For each person, at least one recommendation was indicated as close or distant connection.
Among the connections, the experts identified one recommendation that was close to Benedykt z Koźmina.
Among the distant connections, Sebastian Sierakowski obtained the highest number of distant connections (7), followed by Maciej Karpiga z Miechowa and Benedykt z Koźmina (5).
The graph used in the model was small, so finding closely related individuals may not have occurred very often.
Nevertheless, if such a connection existed, it was found in the model and later confirmed during the evaluation.

The results for \emph{CHExRISH\_FullCAC\_260128} indicate that the model allowed the identification of connections with selected representatives of the University, which were confirmed by experts.
There was an increase in the number of close and distant connections compared to the smaller model, what was expected due to higher ``density'' of the larger graph.

During the evaluation of the \emph{CHExRISH\_Prototype2}, experts did not use the dedicated fields for comments and remarks.
Feedback was only provided during the \emph{CHExRISH\_FullCAC\_260128} evaluation.
Respondents pointed out the lack of certainty due to the lack of additional information about selected individuals (e.g., it is problematic that there were individuals with the same name and from the same town; it was suggested that an additional ID from a database such as CAC would be useful).
Attention was also drawn to the indication of the reason for the algorithm's decision (why this representative was selected by the algorithm).
These elements influenced the difficulty of assessing the connections and their degree: closer and more distant.
Experts also pointed out that close and distant connections, even with the definitions provided (see Sect.~\ref{sec:eval_approach}), remain arbitrary decisions.

\section{Conclusions}
\label{sec:summary}

We presented an end-to-end three stage (embedding, HNSW candidate generation, SPARQL-based semantic filtering) recommendation pipeline for cultural heritage data.
It exemplifies a neuro-symbolic recommender: vector neighbourhoods provide recall, while ontological/SPARQL constraints supply correctness and explanations.
This approach is aligned with current practice in heritage KGs and with sector needs for transparency~\cite{mulholland2024supporting,spillo2024rs}.

The evaluation of the presented prototype shows its usefulness and potential. It also highlights the need for closer cooperation with experts when working on further prototypes.
On the one hand, it is necessary to develop a user interface for the proposed pipeline that will contain identifiers from external databases (unambiguous in comparison to names alone) and explanations that are clear to users.
On the other hand, experts' comments indicated that even simple definitions of correct (close, distant) connections are not entirely clear to experts. Therefore, a workshop with experts is planned to discuss various recommendation scenarios and, as a result, adapt the SPARQL-driven filtering stage to their needs.

Sparse, heterogeneous metadata and broad classes (e.g., \emph{E39\_Actor} conflating individuals and institutions) reduce interpretability---our SPARQL filters and type constraints help, but curatorial data enrichment (temporal spans, roles, linking) remains crucial~\cite{tan2024knowledge}.
With continued data curation, and hybrid modeling (e.g., graph transformers~\cite{yuan2025transformers}), the method can generalize beyond persons to places, events, and artworks.
We also plan to evaluate LLM-assisted, RAG-style relational search over curated subgraphs as a follow-up~\cite{ahola2025sampo}.

\begin{credits}
\subsubsection{\ackname}
We would like to express our sincere gratitude to the experts from the JU Archives, the JU Museum, and the Jagiellonian Library for their time spent evaluating the pipeline.

This publication was funded by a flagship project ``CHExRISH: Cultural Heritage Exploration and Retrieval with Intelligent Systems at Jagiellonian University'' under the Strategic Programme Excellence Initiative at Jagiellonian University.
The research for this publication has been supported by a grant from the Priority Research Area DigiWorld under the Strategic Programme Excellence Initiative at Jagiellonian University.
We gratefully acknowledge Polish high-performance computing infrastructure PLGrid (HPC Center: ACK Cyfronet AGH) for providing computer facilities and support within computational grant no. PLG/2025/018037.

During the preparation of this work the authors used MS Copilot and Writefully in order to improve the readability and language of the manuscript. After using these tools, the authors reviewed and edited the content as needed and take full responsibility for the content of the published article.

\subsubsection{\discintname}
The authors have no competing interests to declare that are
relevant to the content of this article.
\end{credits}

\bibliographystyle{splncs04}
\bibliography{geistbib/culheripub,geistbib/culheriteam}

\begin{thebibliography}{10}
\providecommand{\url}[1]{\texttt{#1}}
\providecommand{\urlprefix}{URL }
\providecommand{\doi}[1]{https://doi.org/#1}

\bibitem{ahola2025sampo}
Ahola, A., Leskinen, P., Rantala, H., Tuominen, J., Hyvönen, E.: Using large
  language models for searching explainable relations in a cloud of cultural
  heritage knowledge graphs: Samposampo as a neuro-symbolic system. In: DHNB
  2026 (October 2025),
  \url{https://seco.cs.aalto.fi/publications/2025/ahola-et-al-samposampo-2025.pdf},
  submitted for review

\bibitem{ali2021pykeen}
Ali, M., Berrendorf, M., Hoyt, C.T., Vermue, L., Sharifzadeh, S., Tresp, V.,
  Lehmann, J.: {PyKEEN} 1.0: A python library for training and evaluating
  knowledge graph embeddings. Journal of Machine Learning Research
  \textbf{22}(82), ~1--6 (2021), \url{http://jmlr.org/papers/v22/20-825.html}

\bibitem{almaaitah_opportunities_2020}
Alma’aitah, W.Z., Talib, A.Z., Osman, M.A.: Opportunities and challenges in
  enhancing access to metadata of cultural heritage collections: a survey.
  Artificial Intelligence Review  \textbf{53}(5),  3621--3646 (June 2020).
  \doi{10.1007/s10462-019-09773-w},
  \url{https://doi.org/10.1007/s10462-019-09773-w}

\bibitem{carriero2019arco}
Carriero, V.A., Gangemi, A., Mancinelli, M.L., Marinucci, L., Nuzzolese, A.G.,
  Presutti, V., Veninata, C.: Arco: The italian cultural heritage knowledge
  graph. In: Ghidini, C., Hartig, O., Maleshkova, M., Sv{\'a}tek, V., Cruz, I.,
  Hogan, A., Song, J., Lefran{\c{c}}ois, M., Gandon, F. (eds.) The Semantic Web
  -- ISWC 2019. pp. 36--52. Springer International Publishing, Cham (2019)

\bibitem{casillo2023context}
Casillo, M., Colace, F., Conte, D., Lombardi, M., Santaniello, D., Valentino,
  C.: Context-aware recommender systems and cultural heritage: a survey.
  Journal of Ambient Intelligence and Humanized Computing  \textbf{14},
  3109--3127 (4 2023). \doi{10.1007/s12652-021-03438-9}

\bibitem{chansanam2023ch}
Chansanam, W., Chaichuay, V., Manorom, P., Kanyacome, S., Sugimoto, S., Huang,
  Y.J., Zhang, G.l., Li, C.: Cultural heritage preservation through
  ontology-based semantic search systems. Conservation Science in Cultural
  Heritage  \textbf{23}(1),  183–196 (2023).
  \doi{10.6092/issn.1973-9494/20048}

\bibitem{chicaiza2021survey}
Chicaiza, J., D{\'{\i}}az, P.V.: A comprehensive survey of knowledge
  graph-based recommender systems: Technologies, development, and
  contributions. Inf.  \textbf{12}(6), ~232 (2021). \doi{10.3390/INFO12060232}

\bibitem{europeana2015metadata}
Dangerfield, M.C., Kalshoven, L.: Report and recommendations from the task
  force on metadata quality. Tech. rep., Europeana (May 2015),
  \url{https://pro.europeana.eu/files/Europeana_Professional/Europeana_Network/metadata-quality-report.pdf}

\bibitem{gaitanou2024ld}
Gaitanou, P., Andreou, I., Sicilia, M.A., Garoufallou, E.: Linked data for
  libraries: Creating a global knowledge space, a systematic literature review.
  Journal of Information Science  \textbf{50},  204--244 (2024).
  \doi{10.1177/01655515221084645}

\bibitem{Haslhofer2018}
Haslhofer, B., Isaac, A., Simon, R.: Knowledge graphs in the libraries and
  digital humanities domain. In: Sakr, S., Zomaya, A. (eds.) Encyclopedia of
  Big Data Technologies, pp.~1--8. Springer International Publishing, Cham
  (2018). \doi{10.1007/978-3-319-63962-8_291-1},
  \url{https://doi.org/10.1007/978-3-319-63962-8\_291-1}

\bibitem{heitmann2010linked}
Heitmann, B., Hayes, C.: Using linked data to build open, collaborative
  recommender systems. In: Linked Data Meets Artificial Intelligence, Papers
  from the 2010 {AAAI} Spring Symposium. {AAAI} (2010),
  \url{http://www.aaai.org/ocs/index.php/SSS/SSS10/paper/view/1067}

\bibitem{hyvonen2019knowledge}
Hyv\"{o}nen, E., Rantala, H.: Knowledge-based relation discovery in cultural
  heritage knowledge graphs. In: Proceedings of the Digital Humanities in the
  Nordic Countries 4th Conference (DHN 2019) (2019)

\bibitem{isaac2012enriching}
Isaac, A.: Case study: Enriching and sharing cultural heritage data in
  europeana. Tech. rep., W3C (June 2012),
  \url{https://www.w3.org/2001/sw/sweo/public/UseCases/Europeana/}

\bibitem{ishchuk2025msc}
Ishchuk, O.: Search and recommendation methods in graph knowledge bases
  describing cultural heritage. Master's thesis, Jagiellonian University
  (2025), supervisor: K. Kutt

\bibitem{luyen2023constraint}
Le, N.L., Abel, M.H., Gouspillou, P.: A constraint-based recommender system via
  rdf knowledge graphs. In: 2023 26th International Conference on Computer
  Supported Cooperative Work in Design (CSCWD). pp. 849--854 (2023).
  \doi{10.1109/CSCWD57460.2023.10152701}

\bibitem{li2023towards}
Li, J., Bikakis, A.: Towards a semantics-based recommendation system for
  cultural heritage collections. Applied Sciences  \textbf{13}, ~8907 (2023).
  \doi{10.3390/app13158907}

\bibitem{mulholland2024supporting}
Mulholland, P., van Kranenburg, P., Carvalho, J., Daga, E.: Supporting the
  end-user curation of cultural heritage knowledge graphs. In: Proceedings of
  the 35th {ACM} Conference on Hypertext and Social Media, {HT} 2024, Poznan,
  Poland, September 10-13, 2024. pp. 35--44. {ACM} (2024).
  \doi{10.1145/3648188.3675132}

\bibitem{cidoc2014primer}
Oldman, D.: The cidoc conceptual reference model (cidoc-crm): Primer. Tech.
  rep., CRM Labs (07 2014),
  \url{http://old.cidoc-crm.org/docs/CRMPrimer_v1.1.pdf}

\bibitem{ric2017primer}
Pitti, D., McCarthy, G., Popovici, B.F.: Records in contexts (ric). an archival
  description draft standard. Tech. rep., ICA Experts Group on Archival
  Description (2017),
  \url{https://www.alaarchivos.org/wp-content/uploads/2018/01/1.-Daniel-V.-Pitti.pdf}

\bibitem{rahman2024faiss}
Rahman, M.S., Rabbi, S.M.E., Rashid, M.M.: Optimizing domain-specific image
  retrieval: {A} benchmark of {FAISS} and annoy with fine-tuned features. CoRR
  \textbf{abs/2412.01555} (2024). \doi{10.48550/ARXIV.2412.01555}

\bibitem{raza2026rs}
Raza, S., Rahman, M., Kamawal, S., Toroghi, A., Raval, A., Navah, F.,
  Kazemeini, A.: A comprehensive review of recommender systems: Transitioning
  from theory to practice. Computer Science Review  \textbf{59},  100849
  (2026). \doi{https://doi.org/10.1016/j.cosrev.2025.100849}

\bibitem{ruotsalo2013smart}
Ruotsalo, T., Haav, K., Stoyanov, A., Roche, S., Fani, E., Deliai, R.,
  M{\"{a}}kel{\"{a}}, E., Kauppinen, T., Hyv{\"{o}}nen, E.: {SMARTMUSEUM:} {A}
  mobile recommender system for the web of data. J. Web Semant.  \textbf{20},
  50--67 (2013). \doi{10.1016/J.WEBSEM.2013.03.001}

\bibitem{schreur2020ld}
Schreur, P.E.: The use of linked data and artificial intelligence as key
  elements in the transformation of technical services. Cataloging \&
  Classification Quarterly  \textbf{58},  473--485 (2020).
  \doi{10.1080/01639374.2020.1772434}

\bibitem{shah2025hnsw}
Shah, H.: A comparative study of hnsw implementations for scalable approximate
  nearest neighbor search  (2025).
  \doi{10.36227/techrxiv.175321947.71782908/v1}

\bibitem{silva2024ch}
Silva, A.L., Terra, A.L.: Cultural heritage on the semantic web: The europeana
  data model. IFLA Journal  \textbf{50},  93--107 (2024).
  \doi{10.1177/03400352231202506}

\bibitem{spillo2024rs}
Spillo, G., Musto, C., de~Gemmis, M., Lops, P., Semeraro, G.: Recommender
  systems based on neuro-symbolic knowledge graph embeddings encoding
  first-order logic rules. User Model. User Adapt. Interact.  \textbf{34}(5),
  2039--2083 (2024). \doi{10.1007/S11257-024-09417-X}

\bibitem{tan2024knowledge}
Tan, M.A.: Knowledge graph construction and refinement for cultural heritage
  digital libraries. In: Taylor, K.L., Zimmermann, A. (eds.) Proceedings of the
  Doctoral Consortium at {ISWC} 2024. {CEUR} Workshop Proceedings, vol.~3884.
  CEUR-WS.org (2024), \url{https://ceur-ws.org/Vol-3884/paper7.pdf}

\bibitem{terras2022dh}
Terras, M.: Digital humanities and digitized cultural heritage. In:
  O’Sullivan, J. (ed.) The Bloomsbury Handbook to the Digital Humanities, pp.
  255--266. Bloomsbury Academic (2022). \doi{10.5040/9781350232143.ch-24}

\bibitem{lvm2025juhmpprototype1}
do~Valle~Miranda, L., Kutt, K., Nalepa, G.J.: {CIDOC-CRM} and the first
  prototype of a semantic portal for the chexrish project. In: Bruns, O.,
  Graciotti, A., Sartini, B., Tietz, T. (eds.) Proceedings of the Second
  International Workshop of Semantic Digital Humanities (SemDH 2025). {CEUR}
  Workshop Proceedings, vol.~4009. CEUR-WS.org (2025),
  \url{https://ceur-ws.org/Vol-4009/paper\_14.pdf}

\bibitem{miranda_wikidata_er_es2c2026}
do~Valle~Miranda, L., Mozolewski, M., Kutt, K., Nalepa, G.J.: A wikidata-based
  workflow for entity reconciliation strategies evaluation: A study on early
  modern polish personal names. In: Proceedings of ESWC 2026 (2026), accepted
  for publication (to appear)

\bibitem{yuan2025transformers}
Yuan, C., Zhao, K., Kuruoglu, E.E., Wang, L., Xu, T., Huang, W., Zhao, D.,
  Cheng, H., Rong, Y.: A survey of graph transformers: Architectures, theories
  and applications. CoRR  \textbf{abs/2502.16533} (2025).
  \doi{10.48550/ARXIV.2502.16533}

\bibitem{zhao2023embedding}
Zhao, X., Wang, M., Zhao, X., Li, J., Zhou, S., Yin, D., Li, Q., Tang, J., Guo,
  R.: Embedding in recommender systems: {A} survey. CoRR
  \textbf{abs/2310.18608} (2023). \doi{10.48550/ARXIV.2310.18608}

\end{thebibliography}

\end{document}